\documentclass[a4paper]{jpconf}
\usepackage{graphicx}
\begin{document}
\title{Recent COMPASS results on Transverse Spin Asymmetries in SIDIS}

\author{Franco Bradamante}

\address{Trieste Section of INFN, Trieste, Italy,\\
on behalf of the COMPASS Collaboration}


\begin{abstract}
After reviewing the most important
COMPASS results on transverse spin effects in SIDIS, 
I will present the recent work we have done
on weighted Sivers asymmetries and 
``transversity induced'' $\Lambda$ polarisation.
Using the high statistics data collected in 2010 on a transversely polarised 
proton target COMPASS has evaluated the transverse momentum weighted Sivers 
asymmetries both in $x$- and in $z$-bins. 
The results are also compared with the standard unweighted asymmetries. 
Using the whole data sample collected over the years on transversely 
polarised deuteron and proton targets COMPASS has measured the transversity 
induced $\Lambda$ polarisation in the reaction 
$\mu N \rightarrow \mu' \Lambda X$. 
Possible future SIDIS COMPASS measurements will 
also be briefly mentioned.
\end{abstract}

\section{Introduction}
In 1996, the COMPASS spectrometer was proposed as a 
"COmmon Muon and Proton Apparatus for Structure and Spectroscopy", 
capable of addressing a large variety of open problems in both hadron 
structure and spectroscopy. 
By now, we have  published a large number of results, and in 
this workshop several COMPASS contributions will cover the various facets 
of the hadron structure, namely the
longitudinal spin structure of the nucleon in semi-inclusive deeply 
inelastic scattering (SIDIS), the 
transverse spin structure of the nucleon in SIDIS,
in exclusive reactions, and in Drell-Yan scattering, 
and
the hadron multiplicities in unpolarised SIDIS \cite{thisw}. 
In this review I will shortly remind the most important results we have 
obtained on the transverse spin structure of the nucleon in SIDIS,
then I will present our most recent work on the 
weighted Sivers asymmetries and on the $\Lambda$ polarisation, and I will 
conclude mentioning the future measurements we are planning.

\section{The TMD PDFs}
The description of the nucleon structure in terms of collinear parton 
distributions functions (PDFs), i. e. distributions depending only on 
Bjorken $x$ and the negative 
squared four-momentum transfer $Q^2$,  has recently been 
generalized to take into account the parton transverse momentum $k_T$. 
A complete picture of the nucleon at leading order needs a total of 
eight transverse momentum dependent (TMD) PDFs. After integration over the 
transverse momentum three of them reduce to the "collinear" PDFs, 
namely the number density, the helicity and the transversity distributions. 
The other five TMD PDFs are sensitive to the direction of the quark 
transverse momentum, they vanish when integrating over $k_T$, and their 
measurement provides important information on the dynamics of the partons 
in the transverse plane in momentum space.

Particularly interesting is the measurement of the SIDIS cross-section when 
the target nucleon is transversely polarised. In this case 8 (5 in case of 
unpolarised lepton beam) different spin-dependent azimuthal modulations 
of the produced hadrons are 
expected, from which invaluable information on the TMD PDFs can be extracted. 
All these target spin dependent azimuthal modulations 
have been investigated by the HERMES and COMPASS 
Collaborations in pioneering experiments.
In particular,
HERMES \cite{Airapetian:2004tw,Airapetian:2009ae} 
and COMPASS \cite{Adolph:2012sn,Adolph:2012sp,Adolph:2014fjw}
are up to now the only SIDIS experiments 
that
have shown that the Sivers function \cite{Sivers:1989cc}, 
the transversity function and the 
Collins function \cite{Collins:1992kk}
are different from zero. 
It is also of interest to remark that out of the 8 possible transverse spin 
azimuthal modulations in the SIDIS cross-section only the Collins and the 
Sivers asymmetries have been found to be different from zero, while the 
others have all been measured to be compatible with zero. 
And that there 
is no clear experimental evidence that the second T-odd TMD PDF, the so 
called Boer-Mulders function, which should show up in the azimuthal 
modulations of the unpolarised SIDIS cross-section, is different from zero. 
It has also to be added that the non-zero results for the asymmetries have 
been obtained scattering leptons on a proton target, while the corresponding 
asymmetries on a deuteron target (as measured by COMPASS) are compatible 
with zero \cite{Alexakhin:2005iw,Ageev:2006da}, 
hinting at a cancellation between the u- and d- quarks 
PDFs.

\section{Transversity induced asymmetries}
\begin{figure*}[tb]
\centering
\includegraphics[width=0.42\textwidth]{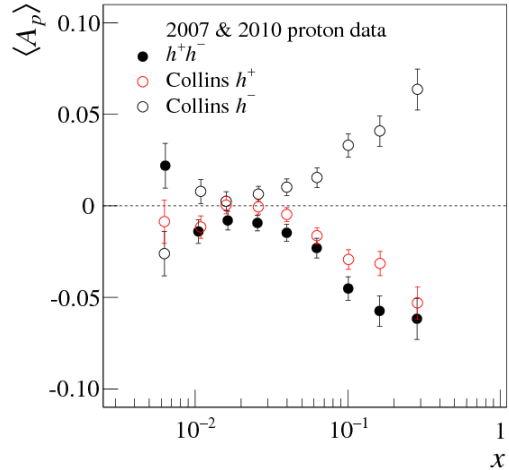}
\caption{Comparison 
of the di-hadron asymmetry with the Collins asymmetry 
from the combined 2007 
and 2010 proton data  \cite{Adolph:2014fjw}.}
\label{fig:1h2h}
\end{figure*}

Some of the most beautiful results of COMPASS are shown in Fig. \ref{fig:1h2h}, 
where the 
Collins asymmetries for positive and negative hadrons are compared to the 
dihadron asymmetry, namely to the azimuthal modulation of the plane containing 
two oppositely charged hadrons. This dihadron asymmetry can be expressed as 
the product of the quark transversity
distribution and a chiral-odd dihadron fragmentation function, which 
survives after integration over the two hadron momenta, and thus can be 
analyzed in the framework of collinear factorization. 
The mirror symmetry of the Collins asymmetry of
positive and negative hadrons and the similarity between the positive hadron 
Collins asymmetry and the dihadron asymmetry suggested us an in-depth 
comparison between the two asymmetries. 
The conclusion of this investigation \cite{Adolph:2015zwe}  
was that both the single hadron and the dihadron transverse-spin dependent 
fragmentation functions are driven by the same elementary mechanism, which 
is very well described in the $^3P_0$ recursive string fragmentation 
model \cite{Artru:2002pua}. 

\section{The Sivers asymmetries}
The Sivers effect was experimentally observed in SIDIS on transversely 
polarised proton targets, first by the HERMES 
Collaboration  \cite{Airapetian:2004tw,Airapetian:2009ae}  
and then by the COMPASS 
Collaboration \cite{Alekseev:2010rw,Adolph:2012sp}, 
while the COMPASS measurements on a transversely 
polarised deuteron target \cite{Alexakhin:2005iw,Ageev:2006da}
gave asymmetries compatible with zero. 
Combined analysis of the proton and deuteron data soon allowed for first 
extractions of the Sivers function for u- and d-quarks
\cite{Efremov:2003tf,Vogelsang:2005cs,Collins:2005ie,Anselmino:2012aa}, which 
turned out to be different from zero, with similar strength and opposite 
sign, a most important result in TMD physics.

In the standard “Amsterdam” notation the Sivers asymmetry can be written 
as 
\begin{eqnarray}
A_{Siv}(x,z) &=&
\frac{\sum_q e_q^2 x f_{1T}^{\perp  \, q}(x) \otimes D_1^q(z)}
     {\sum_q e_q^2 x f_{1}^q(x) \cdot D_1^q(z)}
\label{eq:sa}
\end{eqnarray}
where $x$ is the Bjorken variable,
$z$ is the fraction of the available energy carried  
by the hadron, and
$\otimes$ indicates a convolution over the transverse momenta
of the Sivers function $f_{1T}^{\perp}$ 
and of the fragmentation function $D_1$. 
In all those analyses, due to the presence of the convolution, some functional 
form had to be assumed
for the transverse momentum dependence of the quark distribution 
functions and of the fragmentation
functions.
In most of the analyses these functions were assumed to be Gaussian.
In this case the Sivers asymmetry becomes 
\begin{eqnarray}
A_{Siv,G}(x,z) &=& a_G 
\frac{\sum_q e_q^2 x f_{1T}^{\perp \, (1) \, q}(x) \cdot D_1^q(z)}
     {\sum_q e_q^2 x f_{1}^q(x) \cdot D_1^q(z)} \, .
\label{eq:gsa}
\end{eqnarray}
where 
\begin{eqnarray}
f_{1T}^{\perp (1)}(x) = \int d^2  \vec{k}_T 
\, \frac{k_T^2}{2 M^2} \,  
\, f_{1T}^{\perp}(x, k_T^2) 
\label{k_T_moment}
\end{eqnarray}
and  
$a_G = \sqrt{\pi}M/\sqrt{\langle k_T^2 \rangle _S + 
\langle p^2_{\perp} \rangle /z^2}$
and 
$\vec{p}_{\perp}$ the transverse momentum of the hadron with respect to the quark direction. 
The procedure allows to evaluate from the 
measured Sivers asymmetries the first moment of the Sivers 
function $f_{1T}^{\perp (1)}$, and not the Sivers function itself.

Already twenty years ago an alternative method was proposed 
 \cite{Kotzinian:1995cz,Kotzinian:1997wt,Boer:1997nt} 
to determine $f_{1T}^{\perp (1)}$
without making any assumption on the functional form
neither of the distribution functions nor of the fragmentation functions.
The method consists in measuring asymmetries weighted by the measured
transverse momentum of the hadron $P_T^h$
as described in the next Section.
For some reasons the method was not pursued: 
the only results (still preliminary) came from HERMES  \cite{Gregor:2005qv}. 
Recently, much interest has been dedicated again to the weighted asymmetries 
(see, e.g. \cite{Kang:2012ns}).
In this contribution the first COMPASS results
\cite{Sbrizzai2016,Bradamante:2017yia} 
on the weighted Sivers
asymmetries from the data collected in 2010 with the transversely 
polarised proton target are presented.

\section{$P_T^h/(zM)$ weighted Sivers asymmetries}
To extract the weighted asymmetries,
only the spin-dependent
part of the cross-section has to be weighted,
leaving unweighted the unpolarised cross-section.
In this section the weighting is done with $P_T/(zM)$, where $M$ is the 
nucleon mass.
After some algebra one gets the simple result that
in the numerator of eq. (\ref{eq:sa})
the convolution becomes the product of the first transverse 
moment of the Sivers function and the fragmentation function $D_1$, 
so that the weighted Sivers asymmetry  is
\begin{eqnarray}
A_{Siv}^{w}(x,z) &=&
2\frac{\sum_q e_q^2 x f_{1T}^{\perp \, (1) \, q}(x) \, D_1^q(z)}
     {\sum_q e_q^2 x f_{1}^q(x) \, D_1^q(z)} \,.
\label{eq:wsa}
\end{eqnarray}
Assuming 
u-dominance for positive hadrons produced
on a proton target, the fragmentation function
cancels out and the asymmetry simply becomes
\begin{eqnarray}
A_{Siv}^{w}(x,z) &\simeq&
2\frac{f_{1T}^{\perp \, (1) \, u}(x)}
     {f_{1}^u(x) } \,.
\label{eq:wsag}
\end{eqnarray}

Since the comparison with the “standard” Sivers asymmetry is also important, 
the data production and all the cuts to select the muons and the 
hadrons are the same as for the published data \cite{Adolph:2012sp}.
In particular, the selected phase space is defined by $0.004 < x < 0.7$, 
$Q^2 > 1$ GeV$^2$/c$^2$,  $0.1<y < 0.9$,  $W > 5$ GeV/c$^2$, $P_T > 0.1$ GeV/c, 
and $z > 0.2$.

As shown in \cite{Bradamante:2017yia}, the
 distributions of the weight factor $P_T/(zM)$ in the
different $x$ bins are very similar. Also, the
$P_T/z$ acceptance of the spectrometer is about 60\% and rather flat.  
\begin{figure*}[tb]
\centering
\includegraphics[width=0.48\textwidth]{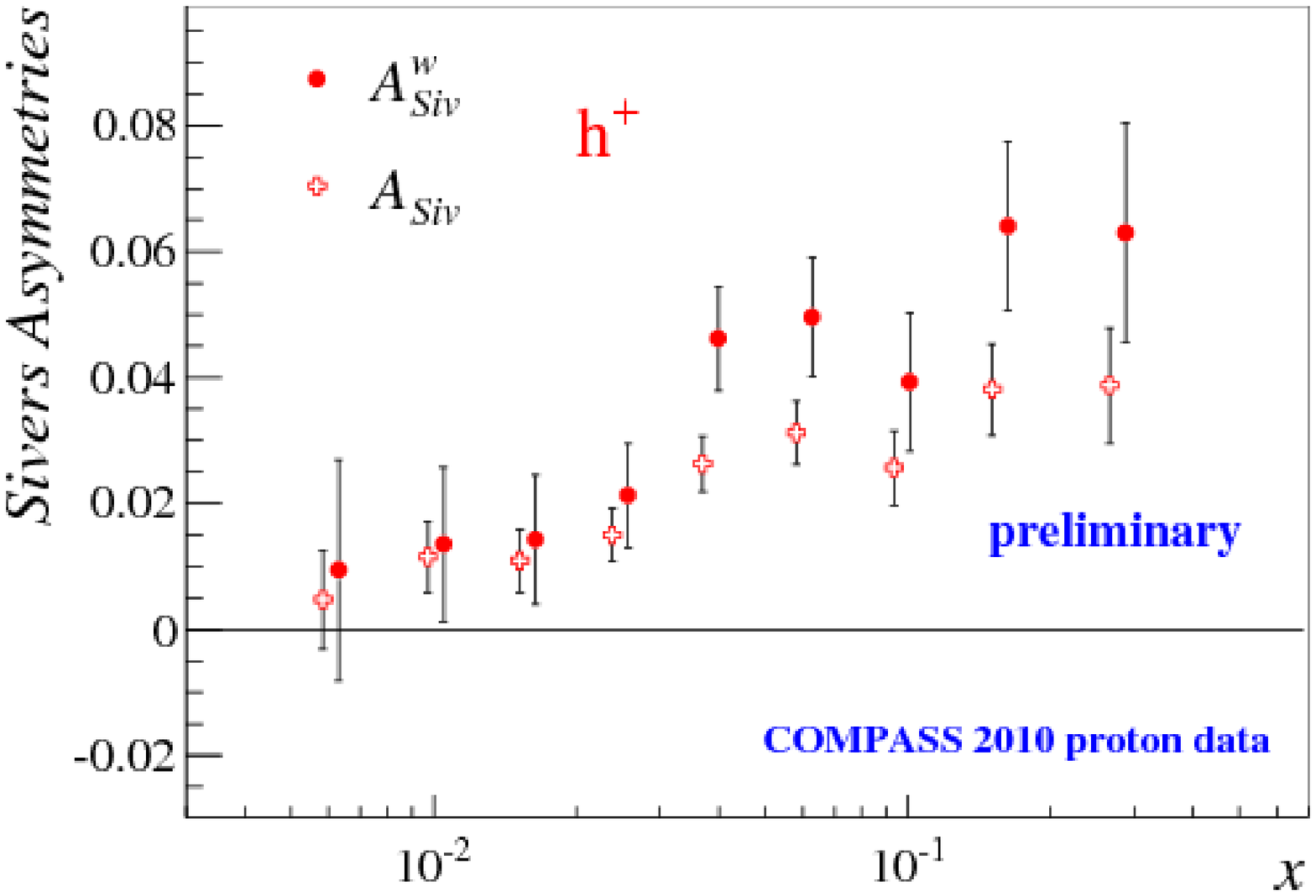}
\includegraphics[width=0.48\textwidth]{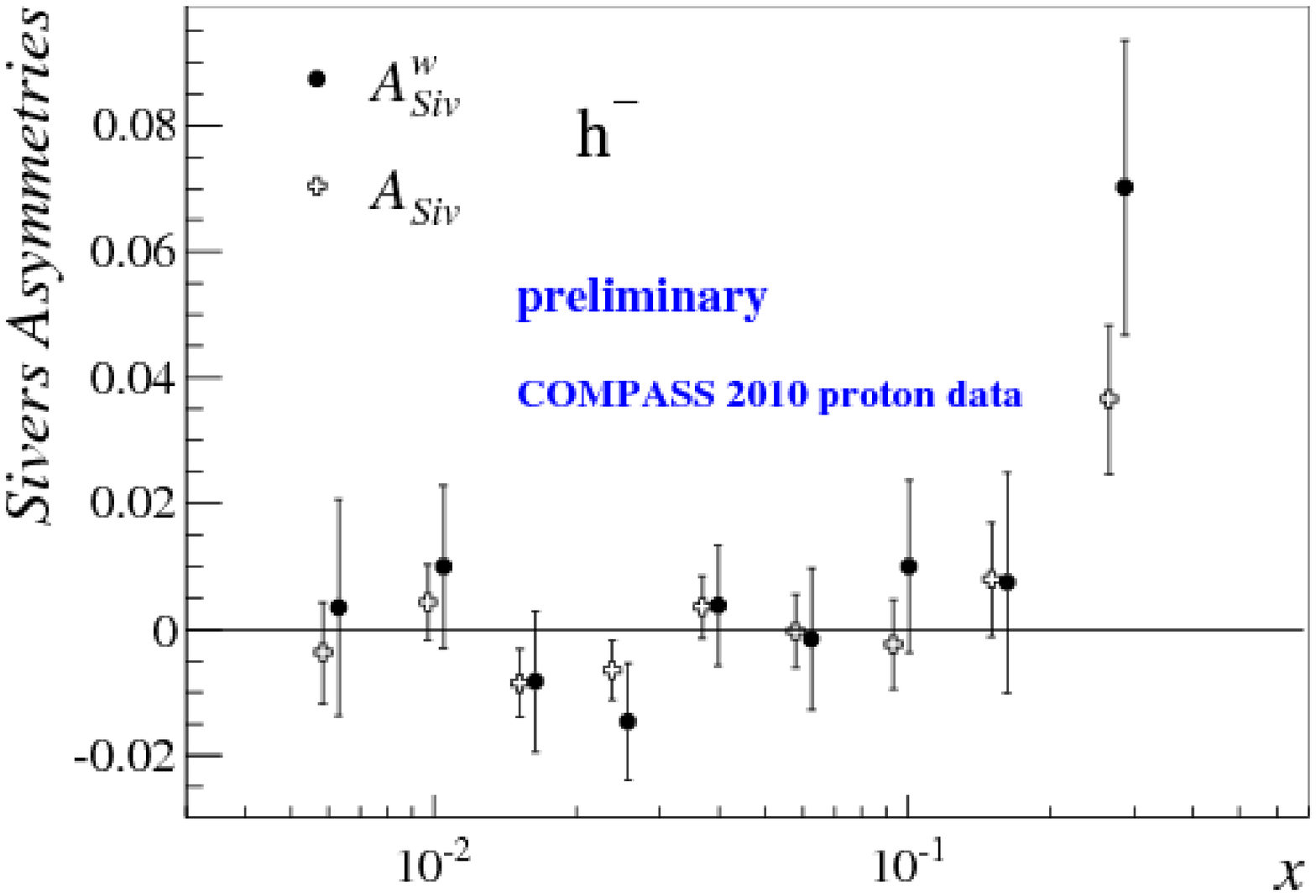}
\hfill
\caption{Full points: $A_{Siv}^{w}$ in the nine $x$ bins for positive 
(left panel) and negative (right panel) hadrons. The open crosses are the
standard Sivers asymmetries  $A_{Siv}$ from Ref. \cite{Adolph:2012sp}.}
\label{fig:results}
\end{figure*}
The results in the case of positive hadrons are given in
Fig. \ref{fig:results}  left, while Fig. \ref{fig:results}
right gives the results for negative hadrons. 
In both cases the “standard” Sivers asymmetries published 
in \cite{Adolph:2012sp}
are also 
plotted for comparison. 
As expected, the trend of the asymmetries is similar
both for positive and negative hadrons.
Assuming u-dominance, the results for positive hadrons 
which are clearly different from zero in particular at large $x$,
where $\langle Q^2 \rangle$ reaches $\sim 20$ GeV$^2$/c$^2$,
constitute the first direct measurement of 
$f_{1T}^{\perp \, (1) \, u}(x) / f_{1}^u(x)$. 

\begin{figure}[tb]
\centering
\includegraphics[width=0.48\textwidth]{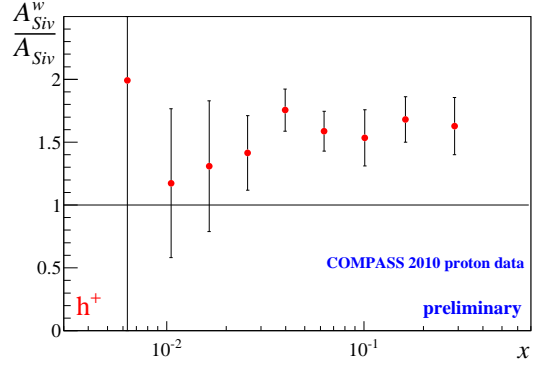}
\caption{The ratios $R^w$ between $A_{Siv}^{w}$ and $A_{Siv}$ in the nine
$x$-bins for positive hadrons.}
\label{fig:ratio_pos}
\end{figure}

Given the similar trend of the weighted asymmetries $A_{Siv}^{w}$ 
and of the standard asymmetries $A_{Siv}$, we have evaluated in each $x$ bin 
their ratios $R^w=A_{Siv}^{w}/A_{Siv}$. 
The values of $R^w$ 
are given in Fig. \ref{fig:ratio_pos} for positive hadrons, which exhibit 
a large Sivers asymmetry. 
In spite of the large statistical uncertainties
the ratios are compatible with a constant value, which is 
rather well determined ($1.6 \pm 0.1$). 
Also, the values in the different $x$ bins agree rather well with those of 
$\langle P_T / (z M) \rangle$, as expected.

\begin{figure*}[tb]
\centering
\includegraphics[width=0.48\textwidth]{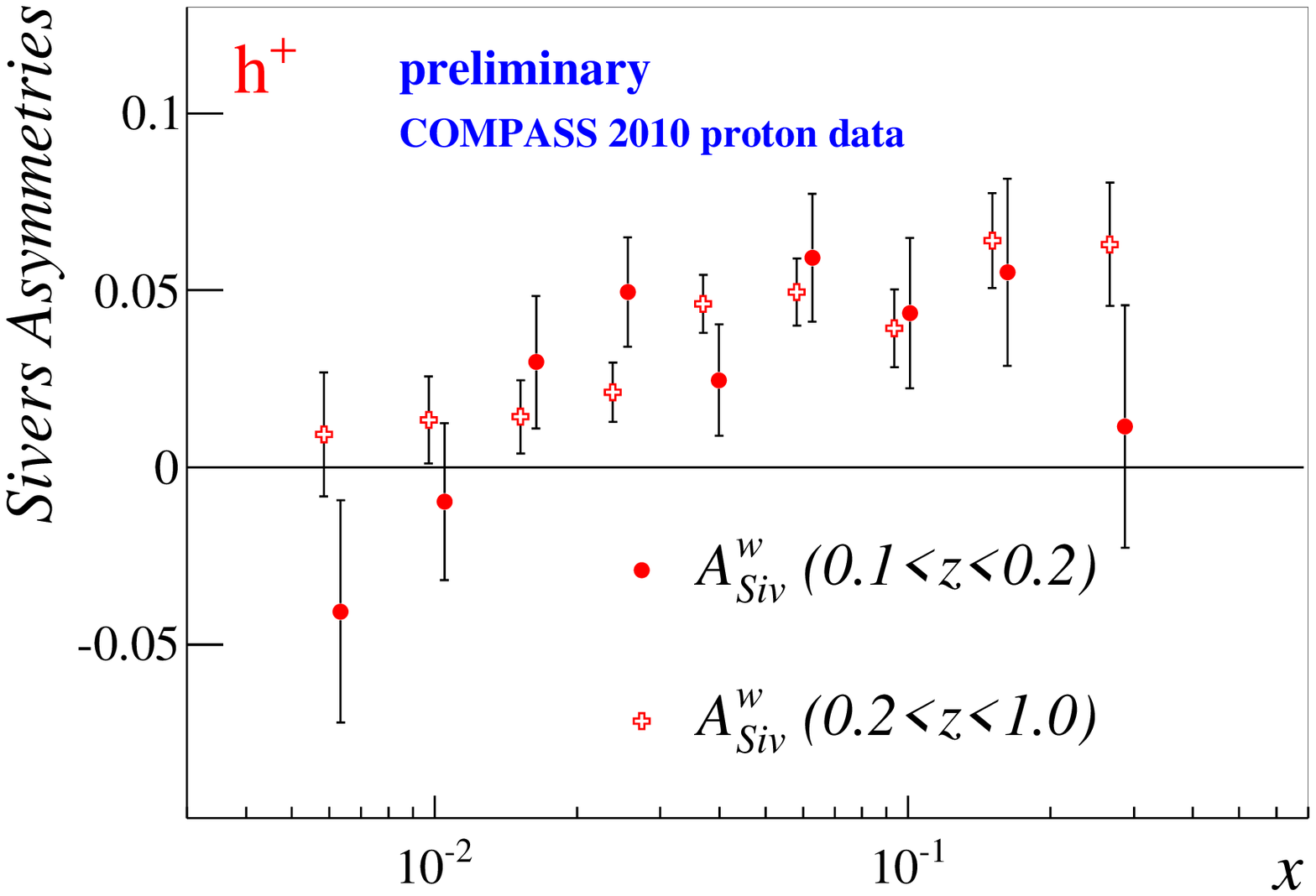}
\includegraphics[width=0.48\textwidth]{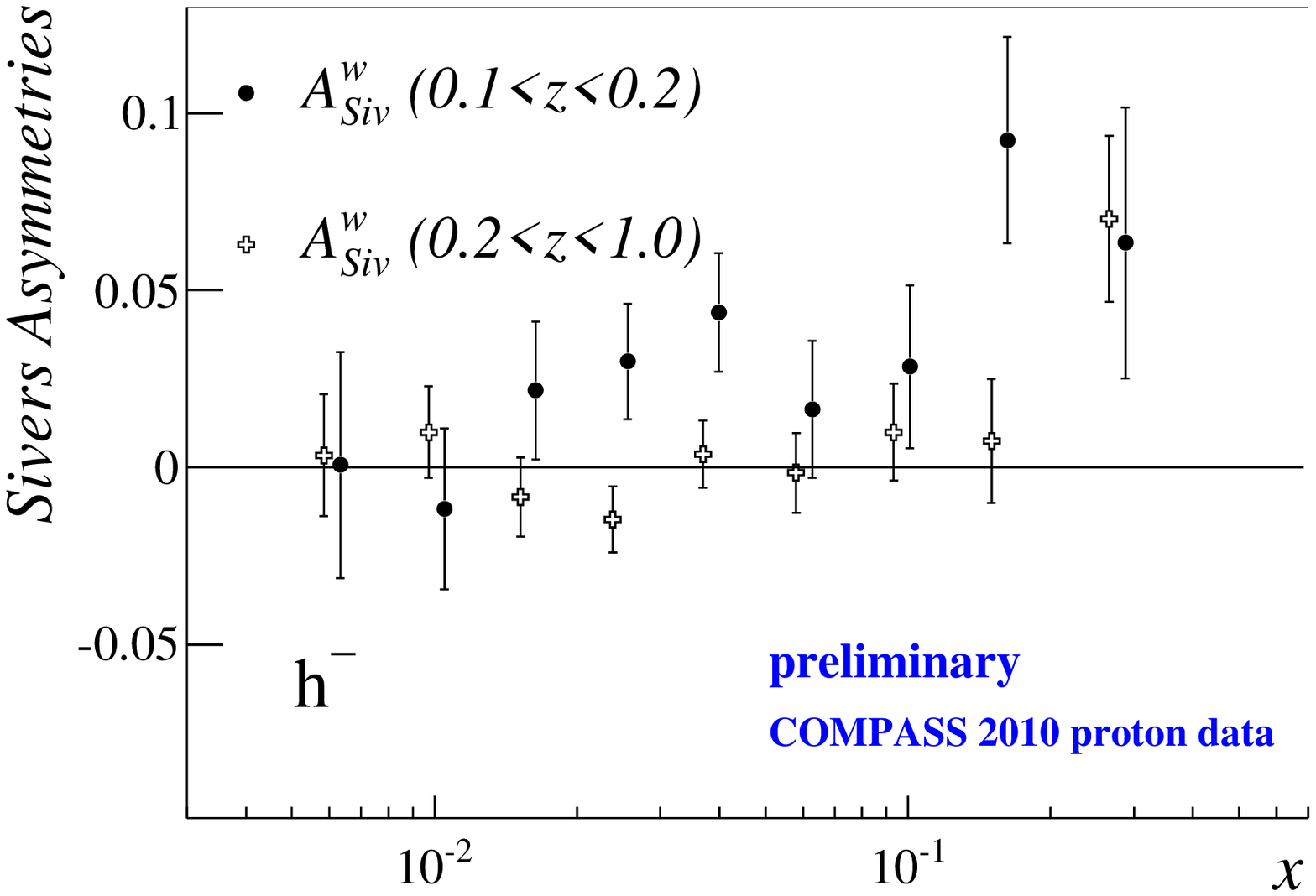}
\caption{Comparison of the weighted asymmetries 
vs. $x$ measured in our "standard" 
$z$ range ($z>0.2$) and the corresponding ones in the range 
($0.1<z<0.2$) for positive (left) and negative (right) hadrons.}
\label{fig:lzhz}
\end{figure*}

In Fig. \ref{fig:lzhz} the weighted Sivers asymmetries measured in our "standard" 
$z$-range ($z>0.2$) are compared with the corresponding ones in the range 
$0.1<z<0.2$. It is interesting to see that the positive hadron asymmetries 
are basically unchanged, hinting again at u-dominance, and supporting the 
idea that factorisation works also at small $z$ in the COMPASS kinematical 
range. 
The negative hadron asymmetries receive contributions from both the u- 
and the d- quark, and become larger and similar to the positive hadron 
asymmetries.
\begin{figure*}[tb]
\centering
\includegraphics[width=0.48\textwidth]{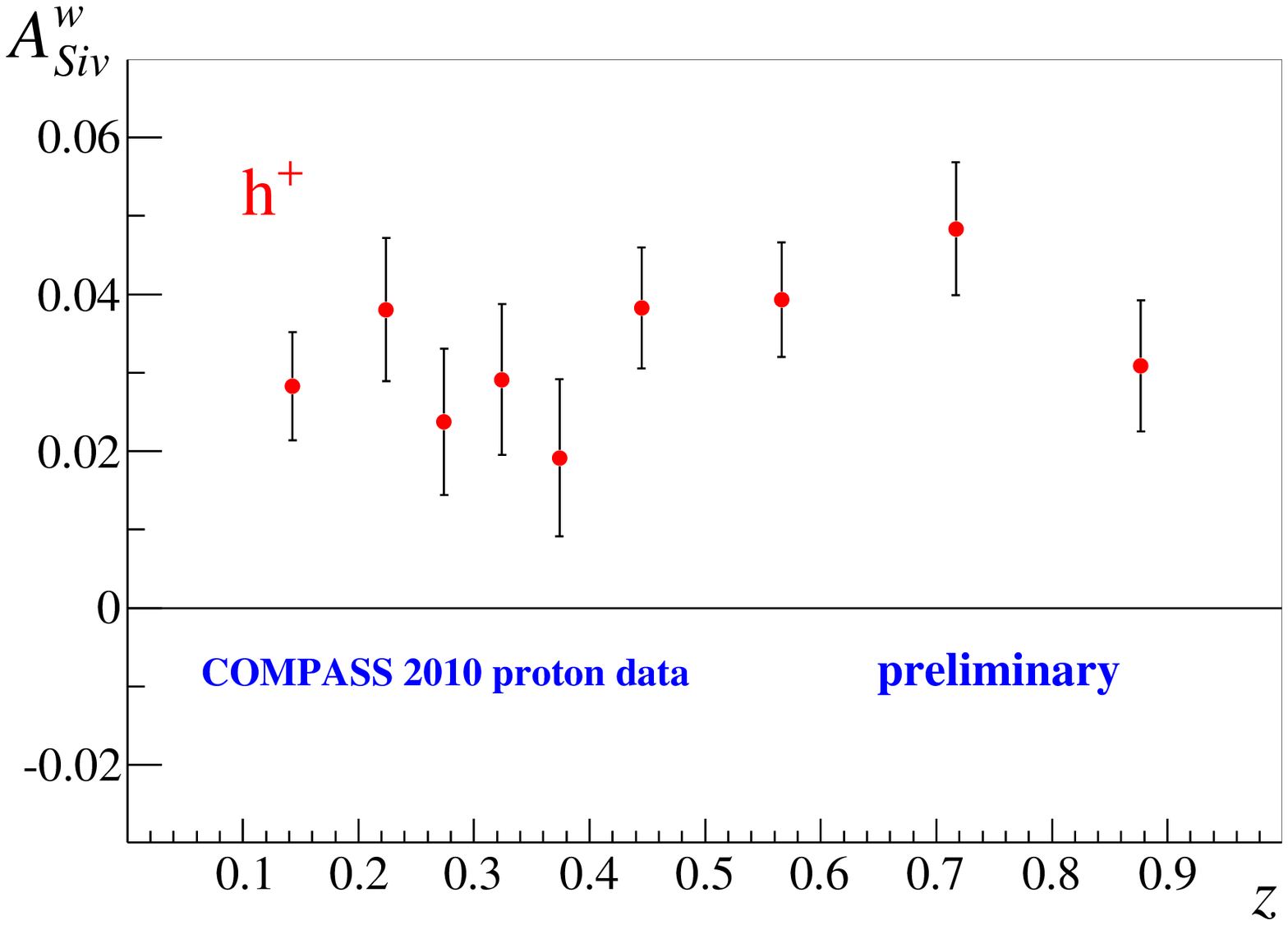}
\includegraphics[width=0.48\textwidth]{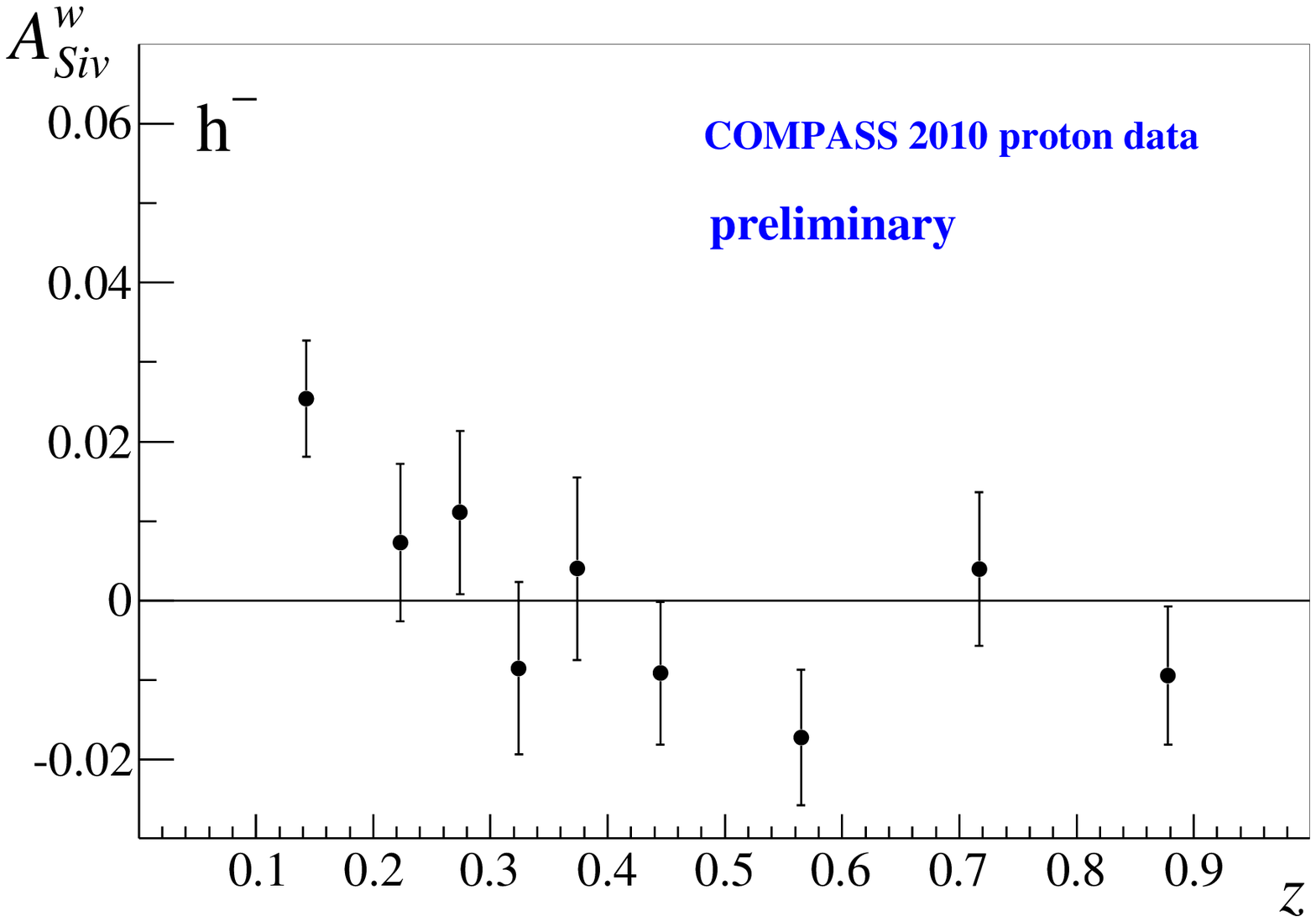}
\caption{Weighted Sivers asymmetries vs $z$ for positive (left) and negative
(right) hadrons.}
\label{fig:zbins}
\end{figure*}
It is of interest to look at $A_{Siv}^{w}$
as a function of $z$, after integration over $x$.
The results in the range $0.1<z<1$ are shown in Fig. \ref{fig:zbins}.
For positive hadrons the values look constant, within the statistical uncertainties, as expected in the case of u dominance.

\section{$P_T^h/M$ weighted Sivers asymmetries}
\begin{figure*}[tb]
\centering
\includegraphics[width=0.48\textwidth]{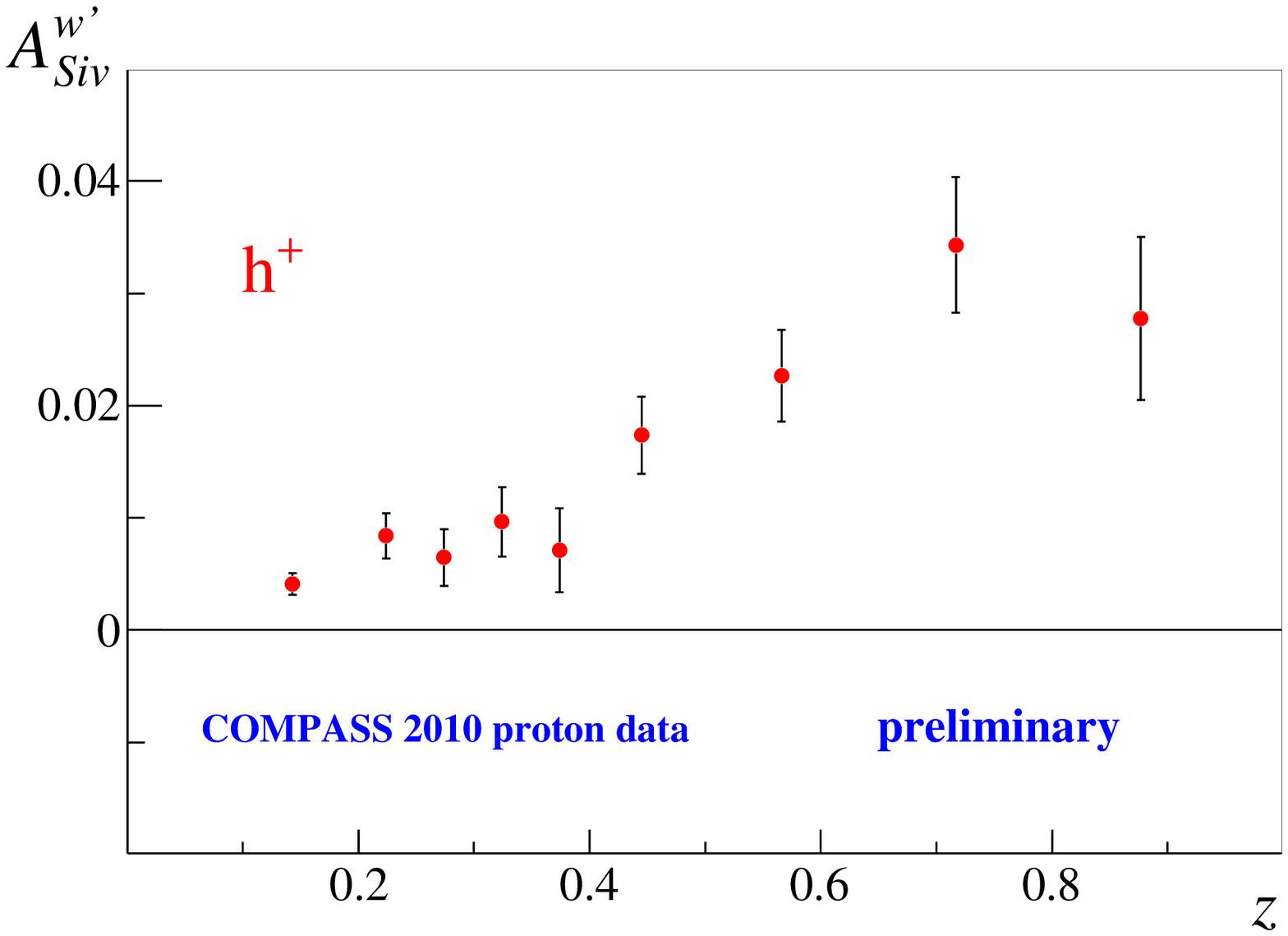}
\includegraphics[width=0.48\textwidth]{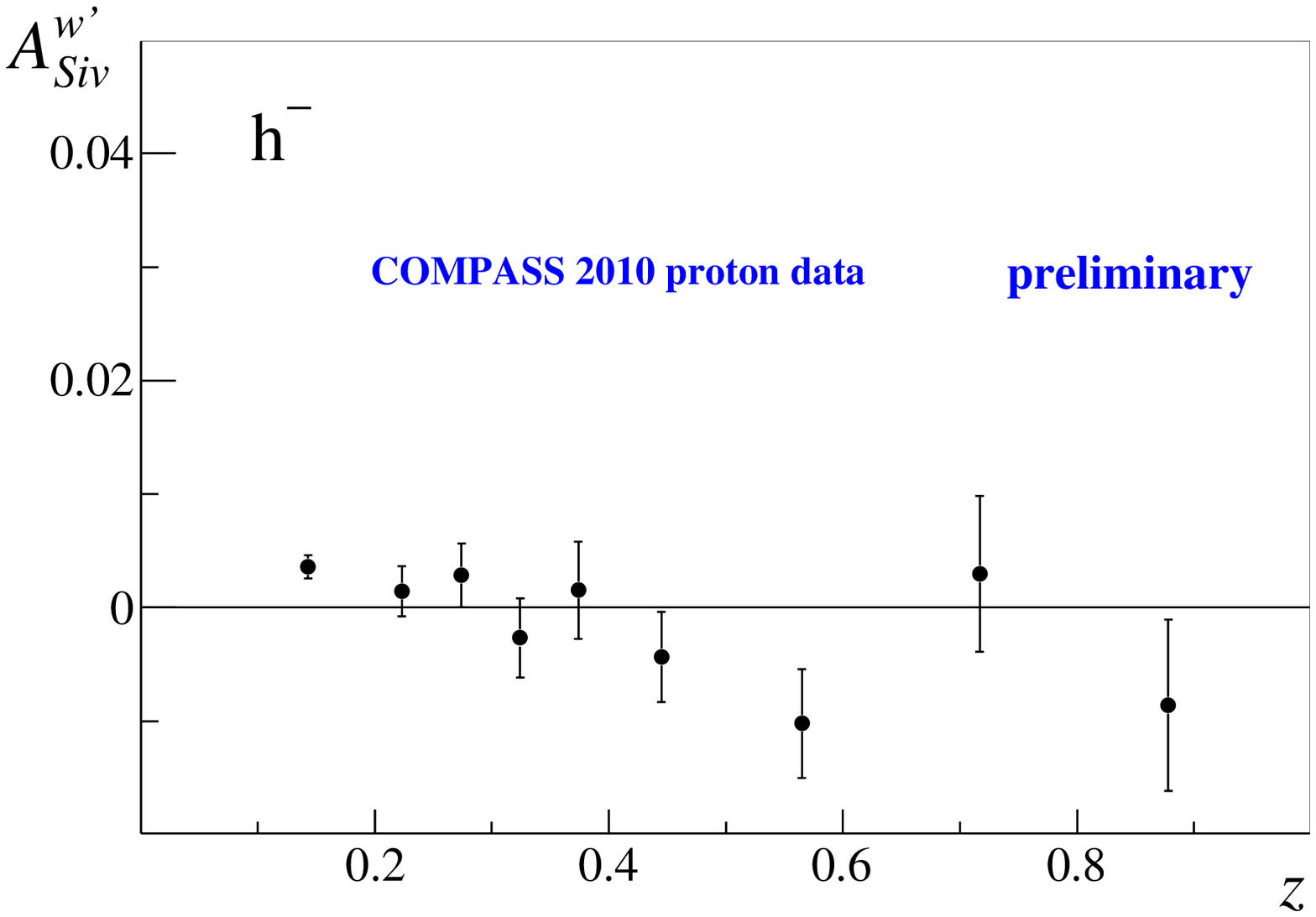}
\caption {$P^h_T/M$ weighted Sivers asymmetries vs $z$ for positive (left) and negative
(right) hadrons.}
\label{fig:xxx}
\end{figure*}

These weighted asymmetries are of interest because their values should 
exhibit a $z$ dependence close to that of the Sivers asymmetries
in the Gaussian model. 
Using as weight $w'=P^h_T/M$, eq. (\ref{eq:wsa}) becomes
\begin{eqnarray}
A_{Siv}^{w'}(x,z) &=& 
2 \frac{\sum_q e_q^2 x f_{1T}^{\perp \, (1) \, q}(x) \, z D_1^q(z)}
     {\sum_q e_q^2 x f_{1}^q(x) \, D_1^q(z)}
\label{eq:wpsa}
\end{eqnarray}
By assuming that the mean value $<P_T^h>$ is independent on $z$ in the measured 
range and that the slopes of the Sivers and the unpolarised PDFs are the
 same, one gets $a_G \simeq \pi Mz/2 <P^h_T>$ \cite{Martin:2017yms} and thus
the Sivers asymmetry with
the Gaussian ansatz in eq.(\ref{eq:gsa}) can be written as 
\begin{eqnarray}
A_{Siv,G}(x,z) &=& 
\frac{\pi M}{2 \langle  P_T^{h} \rangle}
\frac{\sum_q e_q^2 x f_{1T}^{\perp \, (1) \, q}(x) \, z D_1^q(z)}
     {\sum_q e_q^2 x f_{1}^q(x) \, D_1^q(z)} \, ,
\label{eq:gsaa}
\end{eqnarray}
which is similar to the expression $A_{Siv}^{w'}(x,z)$ given in 
eq. (\ref{eq:wpsa}). 
In particular it is 
$R_G^{w'} = A_{Siv}^{w'}/A_{Siv,G} = 4 \langle  P_T^{h} \rangle / \pi M
\simeq 0.7$.
The results for the $x$-integrated asymmetry $ A_{Siv}^{w'}$ 
obtained using $w′= P^h_T/M$ 
are shown in fig. \ref{fig:xxx} as a function of $z$ for positive (left) and 
negative (right) hadrons. 
The values for positive hadrons are in qualitative agreement with 
the expectation:
\begin{eqnarray}
A_{Siv}^{w'}(z) &=& 
2 \, z \, \frac{\int_{\Omega_x} dx C(x) x f_{1T}^{\perp \, (1) \, u}(x)}
     {\int_{\Omega_x} dx C(x) x f_{1}^u(x)}
\label{eq:wsa_zu}
\end{eqnarray}

\section{$\Lambda$ polarisation}

The measurement of the transverse polarisation of $\Lambda$ hyperons 
produced in SIDIS off transversely polarised nucleons has always been 
indicated as a promising channel to access transversity \cite{Artru:1990wq}. 
The basic idea is that, if transversity is different from zero, the polarisation of the fragmenting quark is transferred to the 
$\Lambda$ according to the transversely polarised fragmentation function 
$H_1^{\Lambda  /q}$, so that the $\Lambda$ is polarised.
Now we know that both the u- and the d- quark transversity are different 
from zero, so it is possible to check whether there is any transfer 
of polarisation from them to the lambda, and obtain information on
$H_1^{\Lambda /q}$.
Summing over all quark species, we get:
\begin{equation}
\vec{P}_{\Lambda} =\frac{\sum_q e_q^2 h_1^q(x) H_1^{\Lambda/q}(z)}{\sum_q e_q^2 f_1^q(x) D_1^{\Lambda/q}(z)}D_{NN}R\vec{P}_N 
\end{equation}
where $R$ is the rotation about the normal to the scattering plane 
which brings the initial proton momentum along the $\Lambda$ momentum
 \cite{Artru:1993fn}.
$\Lambda$ hyperons are a natural choice to investigate such transversity 
transmitted polarisation: their weak decay channel 
$\Lambda \rightarrow p \pi^-$  
is in fact self-analysing, that is, the polarisation is revealed through 
the angular asymmetry of the decay proton:
\begin{equation}
\frac{dN}{dcos\theta} \propto 1+ \alpha P_{\Lambda}\cos\theta
\end{equation}
where $\alpha$ is the $\Lambda$ 
weak decay parameter and $\theta$ is the angle between the proton momentum 
and the polarisation vector, assumed to be coincident with that of 
the outgoing quark. 

Studies of $\Lambda$ transversity transmitted polarisation were already 
carried out 
in COMPASS using 2002-2004 deuteron data and 2007 proton data, but the 
results have never been published. 
Now the analysis has been resumed and extended to the proton data collected 
in 2010. 
Since some aspects of the analysis are new, the 2007 data have also been 
re-analysed and results are presented here for the whole set of proton data. 
The $\bar{\Lambda}$ 
polarisation has also been extracted from the complete proton data set.

\begin{figure}[!h]
  \centering
    \includegraphics[width=0.48\textwidth]{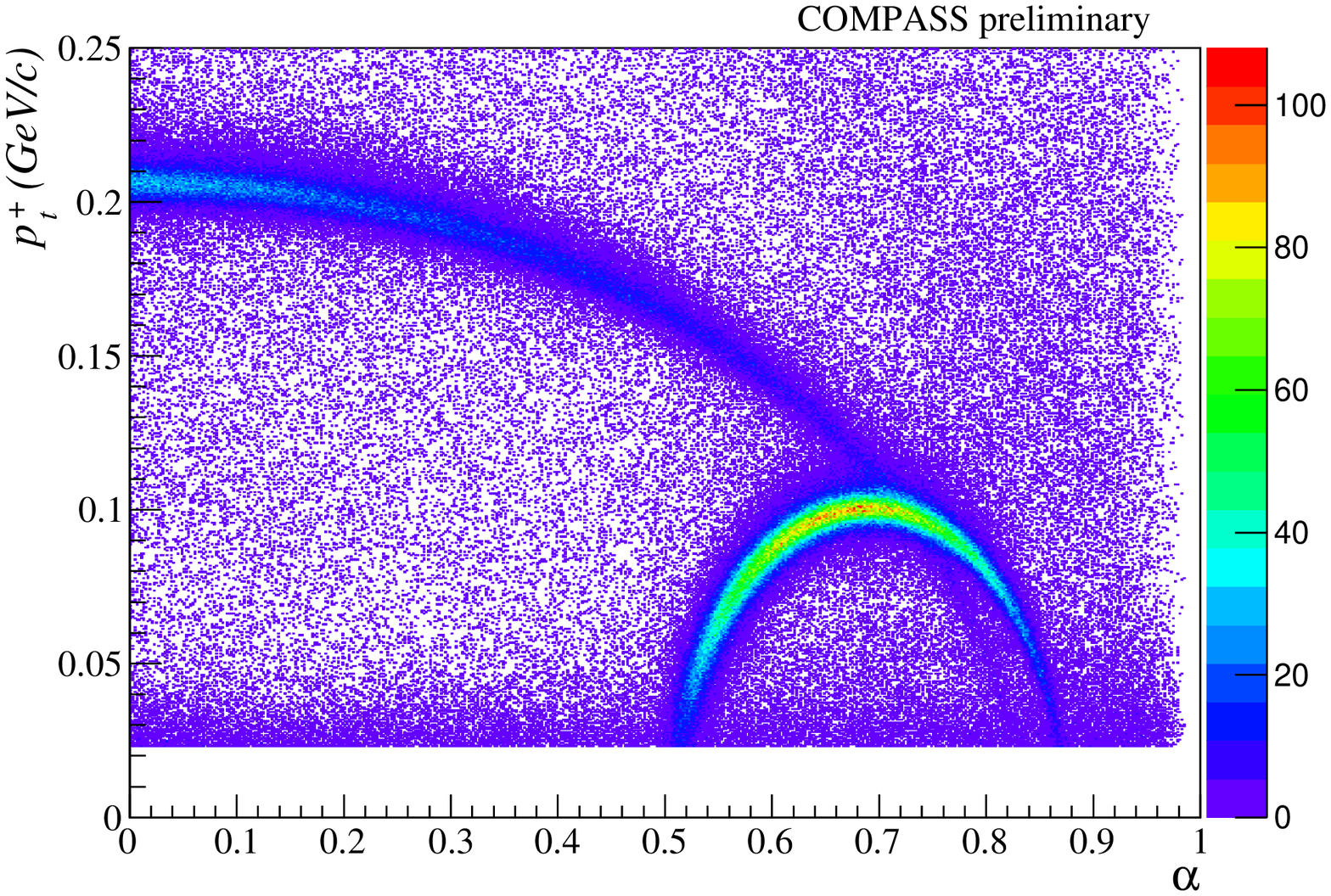}
    \includegraphics[width=0.48\textwidth]{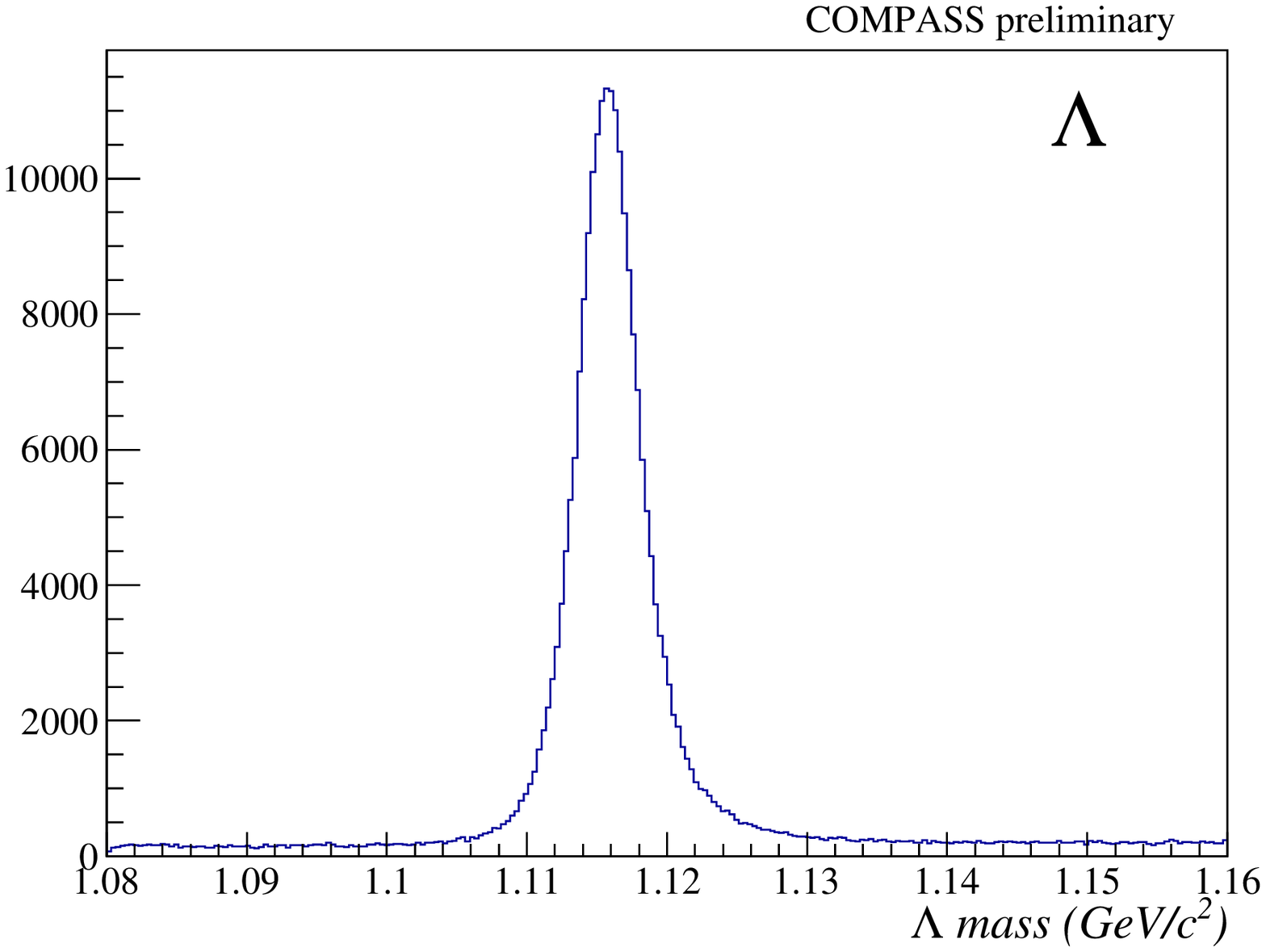} \hfill
    \caption{Armenteros plot after all but mass cuts and invariant mass spectrum for $\Lambda$ candidates.  }\label{fig:lambda1}
\end{figure}

\begin{figure}[!h]
  \centering
    \includegraphics[width=0.48\textwidth]{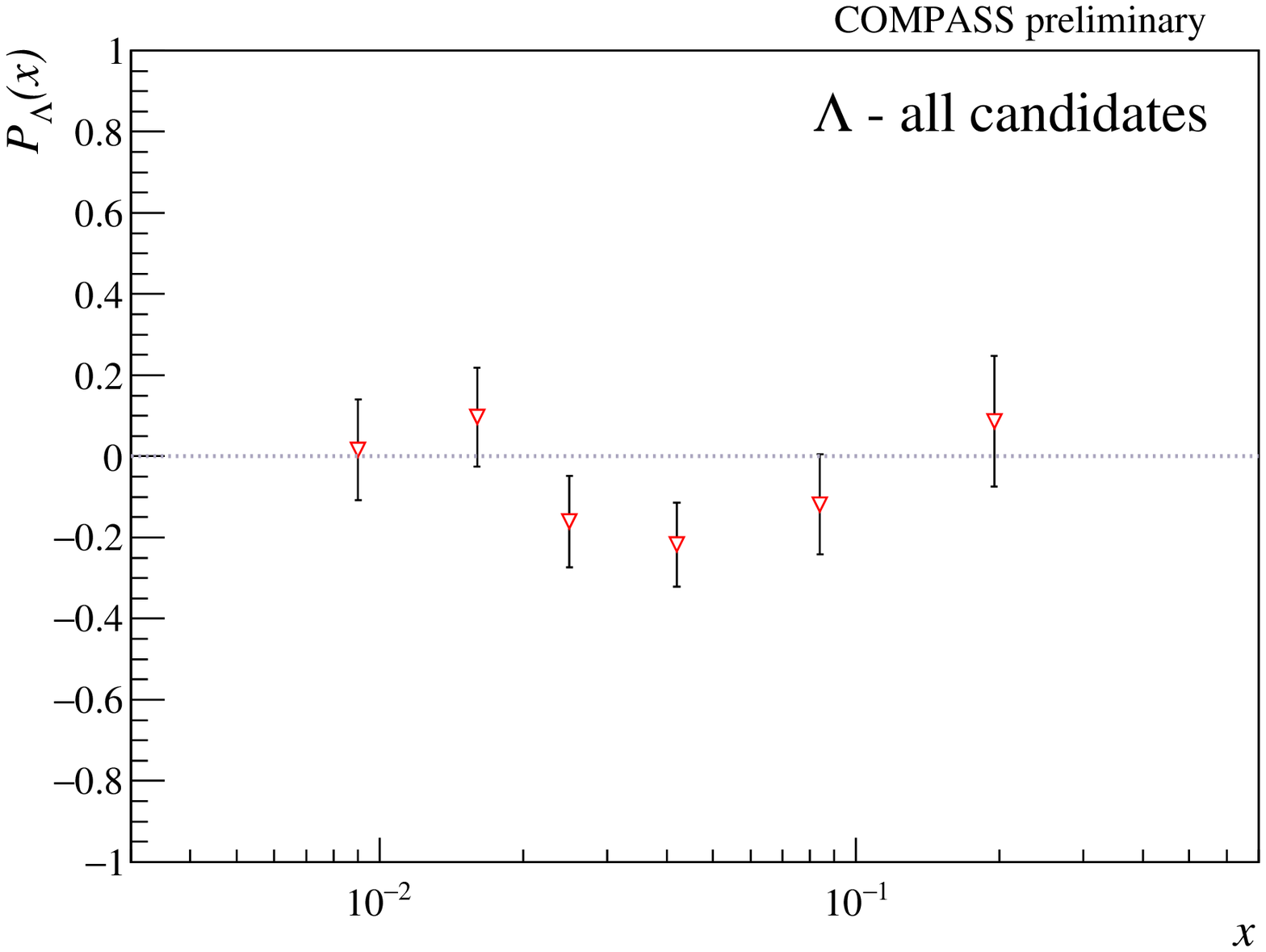}
    \includegraphics[width=0.48\textwidth]{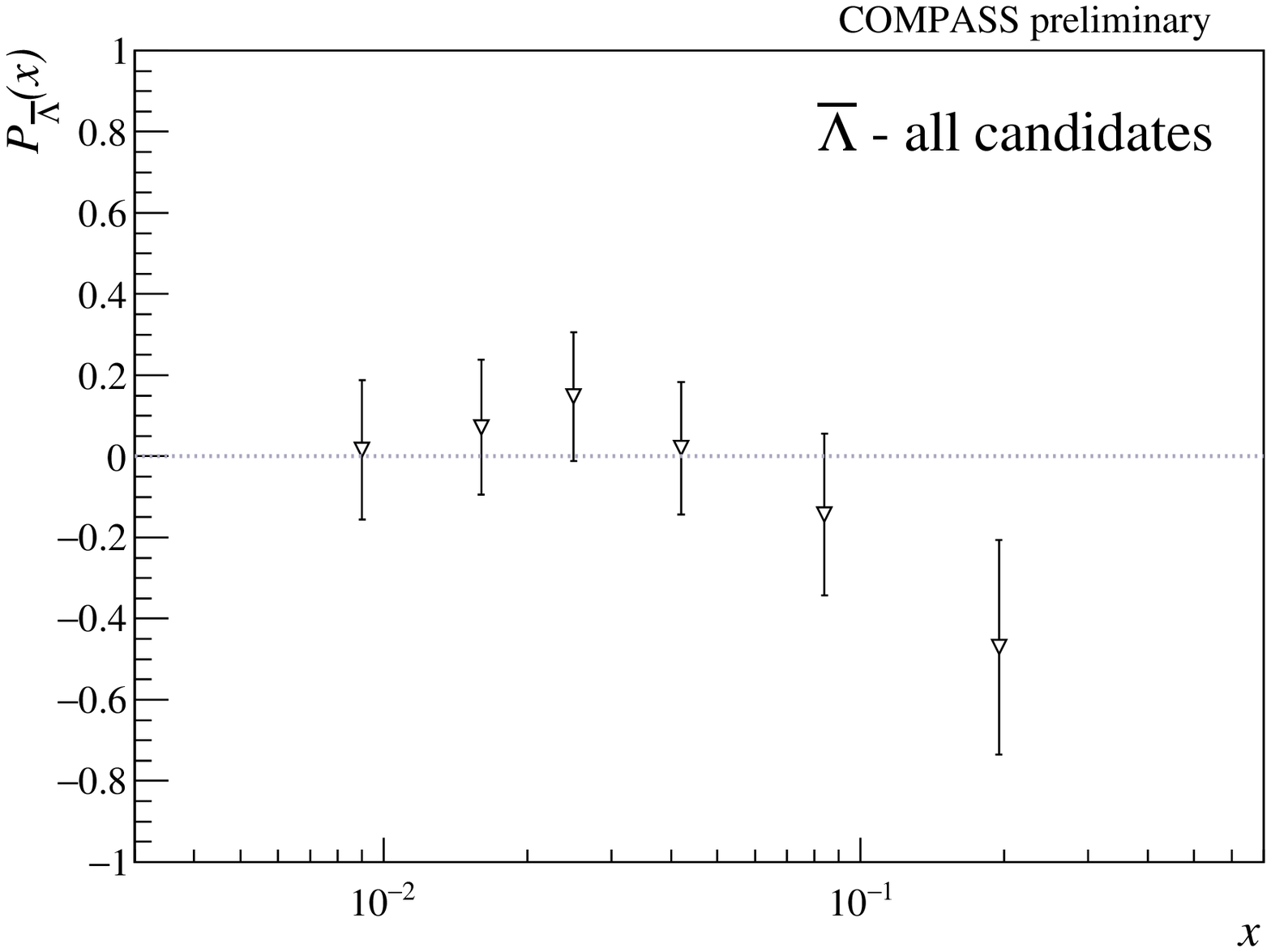}
     \caption{Polarisations for all $\Lambda$ (left) and $\bar{\Lambda}$ (right) candidates as function of $x$. } \label{fig:lambda2}
\end{figure}

In Fig. \ref{fig:lambda1} both the Armenteros-Podolanski plots and the mass 
distribution for $\Lambda$ candidates from 2010 data
are shown after all cuts. 
The results for the polarisation are given in Fig \ref{fig:lambda2}
as a function of $x$. 
All the values are compatible with zero.
                                                   
\section{Future SIDIS measurements}
The Collaboration has submitted to the CERN SPS Committee a proposal for an 
extension of the present COMPASS-II experimental programme to perform in the 
years 2021 and 2022 two new measurements,  muon  SIDIS on a transversely 
polarised deuteron target and elastic muon proton scattering \cite{addc2}.
The first program is the “missing piece” in the COMPASS data sets on 
transverse target spin orientations. While in 2010 a good statistics sample 
of SIDIS data on transversely polarised proton target was collected, in the 
years 2002, 2003 and 2004, due to the late delivery of the new polarised 
target magnet and to the shortness of the periods of data taking, we provided 
only a marginal (albeit unique) data set for the isoscalar deuteron target. 
With one additional year of deuteron target data taking accurate flavor 
separation for the new PDFs will be possible, improving in particular our 
knowledge of the d-quark distribution functions. The second measurement 
hopefully will provide a decisive contribution to the solution of the 
so-called “proton radius puzzle”.\\

\section*{References}
\bibliography{fb_dspin17_bib}

\end{document}